# The orbit of the Chelyabinsk event impactor as reconstructed from amateur and public footage

Jorge I. Zuluaga[a], Ignacio Ferrín[a], Stefan Geens[b]

[a]*Instituto de Física – FCEN, Universidad de Antioquia, Clle. 67 No. 53-108, Medellín, Colombia*
[b]*Ogle Earth, c/o Alpen, Regeringsgatan 87, Stockholm, Sweden*

**Abstract**

A ballistic reconstruction of a meteoroid orbit can be made if enough information is available about its trajectory inside the atmosphere. A few methods have been devised in the past and used in several cases to trace back the origin of small impactors. On February 15, 2013, a medium-sized meteoroid hit the atmosphere in the Chelyabinsk region of Russia, causing damage in several large cities. The incident, the largest registered since the Tunguska event, was witnessed by many thousands and recorded by hundreds of amateur and public video recording systems. The amount and quality of the information gathered by those systems is sufficient to attempt a reconstruction of the trajectory of the impactor body in the atmosphere, and from this the orbit of the body with respect to the Sun. Using amateur and public footage taken in four different places close to the event, we have determined precisely the properties of the entrance trajectory and the orbit of the Chelyabinsk event impactor. We found that the object entered the atmosphere at a velocity ranging from 16.0 to 17.4 km/s in a grazing trajectory, almost directly from the east, with an azimuth of velocity vector of 285º, and with an elevation of 15.8º with respect to the local horizon. The orbit that best fits the observations has, at a 95% confidence level, a semi-major axis $a$ = 1.26±0.05 AU, eccentricity e = 0.44±0.03, argument of perihelion ω = 95.5º±2º and longitude of ascending node Ω = 326.5º±0.3º. Using these properties the object can be classified as belonging to the Apollo family of Near Earth Asteroids. The absolute magnitude of the meteoroid was H= 25.8, well below the threshold for its detection and identification as a Potential Hazardous Asteroid (PHA). This result would imply that present efforts intended to detect and characterize PHAs are incomplete and are missing approximately half the objects able to impact our planet and cause local damage.

## 1. Introduction

There is an increasing level of awareness about the potential risks posed by the impact of medium-sized to large Near Earth Asteroids (NEAs). Global efforts have been made to detect, characterize and predict the trajectories of the several thousand asteroids that have a non-negligible probability of impacting our planet and causing damage to highly populated areas. And yet there remain unpredictable events such as the one witnessed in the Chelyabinsk region of Russia, where a small

*Corresponding author at*: Instituto de Física – FCEN, Universidad de Antioquia, Clle. 67 No. 53-108, Medellín, Colombia.
*Email addresses*: jzuluaga@fisica.udea.edu.co (Jorge I. Zuluaga), ferrin@fisica.udea.edu.co (Ignacio Ferrin), stefan.geens@gmail.com (Stefan Geens)
*Abreviations*: HA, Harmful Asteroids; ChEA, Chelyabinsk Event Asteroid

asteroid with a diameter estimated at 17-20 m hit the atmosphere, producing a super bolide and an explosion with an energy of around 440 kton (Yeomans & Chodas, 2013).

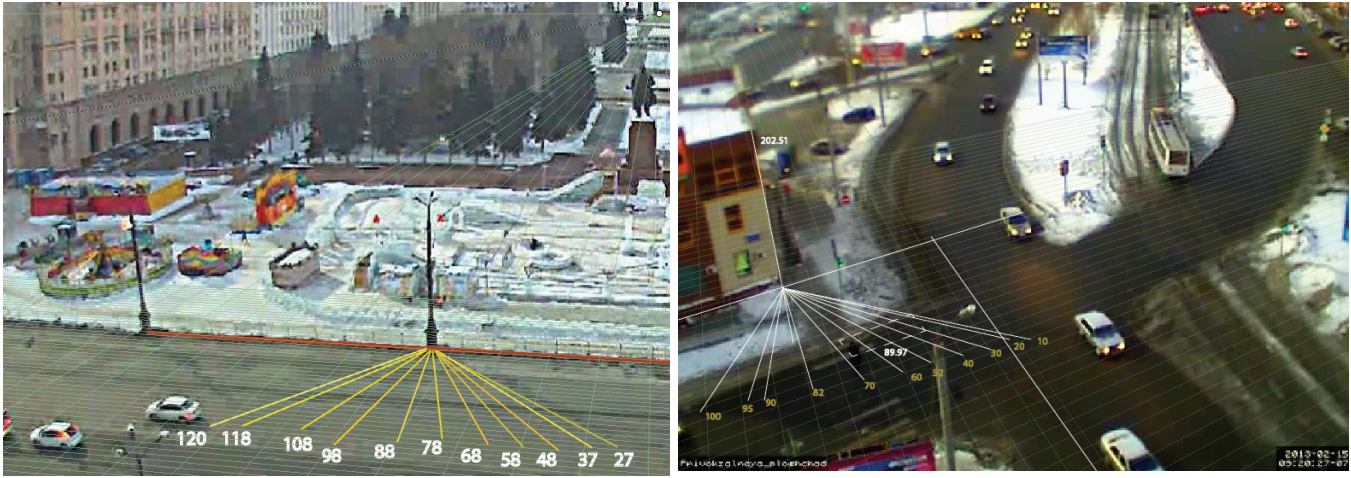

**Figure 1. Footage from two of the vantages points used in this work: the Revolution Square in Chelyabinsk (left panel and Privokzalnaya Square in the same city**. Lines show the direction and length of the shadow at different times in the video. The numbers indicate an arbitrary label for the snapshots where the shadows where measured.

The Chelyabinsk event happened at 09:20 local YEKT time (UTC+6 hours), just after sunrise and at a time when over a million people were awake and potential witnesses to the display. Hundreds of amateur cameras, mobile phones, security and public cameras and even professional recording devices registered the development of the atmospheric event from its first appearance up to the powerful explosion and fragmentation of the parent body at a relatively low altitude. This event, the largest recorded since the Tunguska impact (Andreev, 1990; Sekanina, 1998; Gasperini, et al., 2007) has been historic in several aspects. First, it happened over a region enclosing populated areas (there are four cities near the impact region with a combined population of 1.5 million). Second, it happened close to a peak commuting time, when many people were outdoors. Third, it happened in an era when devices capable of recording video are cheap and ubiquitous, allowing pervasive observations from many different local vantage points. And finally, it happened in the information age, when anybody is able to share almost instantaneously their experiences through social networks. All these conditions ensured that this unexpected but historic natural event would become one of the best documented ever.

The reconstruction of the impactor trajectory in the case of super bolides has become customary (see e.g. Balabh et al. 1978, Trigo-Rodríguez et al. 2010). Several observing networks have been installed around the world to detect and facilitate the ballistic reconstruction of the parent body (see e.g. Caplecha, 1987; Trigo-Rodríguez et al. 2006). Citizen has been also involved in these efforts in the past (Huziak & Sarty, 1994). Despite these efforts, only several orbits have been successfully determined, mainly due to the lack of enough triangulation information (see e.g. Trigo-Rodríguez et al. 2007). Although no known meteor network registered the Chelyabinsk super bolide, a large number of amateur and public video recordings acted as an ad-hoc sensor network, able to potentially provide information for a successful reconstruction of the trajectory through the atmosphere. Challenges in using this method include the relative scarcity of this ad-hoc network, the variable quality of the footage, and the lack of synchronization between the time recording systems. However, the pervasive presence of GPS devices and free public access to accurately positioned satellite imagery within the context of virtual globes such as Google Earth allow the pinpointing of the precise geographic locations of these recording devices.

## 2. Methods

After checking almost a hundred amateur videos and recordings taken by security and public cameras, we selected four videos for their high quality and for the availability of precise location information, both elements essential to a successful triangulation of the impactor trajectory. The videos were recorded at (1) Revolution Square in downtown Chelyabinsk (55.16080°N, 61.40249°E), (2) Privokzalnaya square just 2 km to the south (55.14392°N, 61.41421°E), (3) the marketplace of Korkino, a city 30 km to the south of Chelyabinsk (54.89093°N, 61.39956°E) and (4) the Central Square of Kamensk-Uralsky (56.41500°N, 61.91858°E), a city 150 km to the north of the impact. For additional information about the vantage points, see Appenix A. The base of the triangulation covers a longitudinal distance of around 170 km, which is of the same order of magnitude as the length of the impactor trajectory.

The highest quality videos were analyzed frame by frame during the 5-second duration of the brightest part of the event. Using a method originally devised by one of us (S. Geens) we measured the shadows cast by objects in the scene and estimated from them the elevation and azimuth relative to reference directions of the fireball as a function of time (see figure 1). This method was used in three of the four recordings (Revolution Square, Privokzalnaya Square and Korkino), where cameras were pointing to the floor and no images of the fireball in the sky were available (in the Korkino market place a brief appearance of the bolide is available but at very low quality). In the last recording (Kamensk-Uralsky central square) direct images of the bolide were analyzed, providing additional information about the trajectory. The azimuths of reference directions, distances and sizes of the familiar objects in the footage were obtained from information provided by Google Earth and some archived images of the places used in the triangulation. In order to accurately measure lengths from a two dimensional image, we have corrected for perspective (see perspective grid used in figure 1). Since no information is presently available about the physical properties of the recording devices, no corrections for optical deformations were applied to the measurements. Figure 1 shows snapshots of the measurements performed on the footage at two of the vantage points used in this reconstruction.

We assign errors to azimuths and elevations according to the uncertainty of the determination of, for example, the tip of the shadow or the center of the bolide image in the Kamensk-Uralsky footage. Accordingly, the assumption that the errors for these quantities are normal is reasonable.

One of the trickiest parts of a triangulation procedure is the precise synchronization of the observations performed from different vantage points (see e.g. Borovicka, 1990; Green, 2010). Only two of the videos (those at Revolution Square and Privokzalnaya Square) had proper synchronized time stamps. Those sites were however the closest together and a synchronized triangulation of the trajectory using these observations is affected by uncertainties. In order to avoid the requirement of synchronization, we devised an asynchronous method we called the *altazimuth-footprint method*.

In the method, as many observed elevations and azimuths as possible are measured at a given vantage point and compared with the *altazimuth footprint* of a test trajectory calculated with one of our orbits (see figure 2). For each observation point and at each vantage location we compute the elevation expected for a given test trajectory at the measured azimuth $A_o$ of that point. The expected elevations $h_t$ were then compared with the observed ones $h_o$. For that purpose we computed the least-square statistics:

$$\chi^2(\mathbb{T}) = \sum_{j=1}^{n_v} \sum_{i=1}^{n_j} \frac{[h_{oji} - h_t(A_{oi}; \mathbb{T}, \mathbb{V}_j)]^2}{\Delta h_{oji}}$$

Here the index $j$ runs over the $n_v = 4$ vantage points (represented by $\mathbb{V}_j$) and the index $i$ over the $n_j$ observations performed at that $j$-th vantage point. $\Delta h_{oji}$ is the error of the $i$th observed elevation at the $j$th vantage point. $\mathbb{T}$ represents the test trajectory.

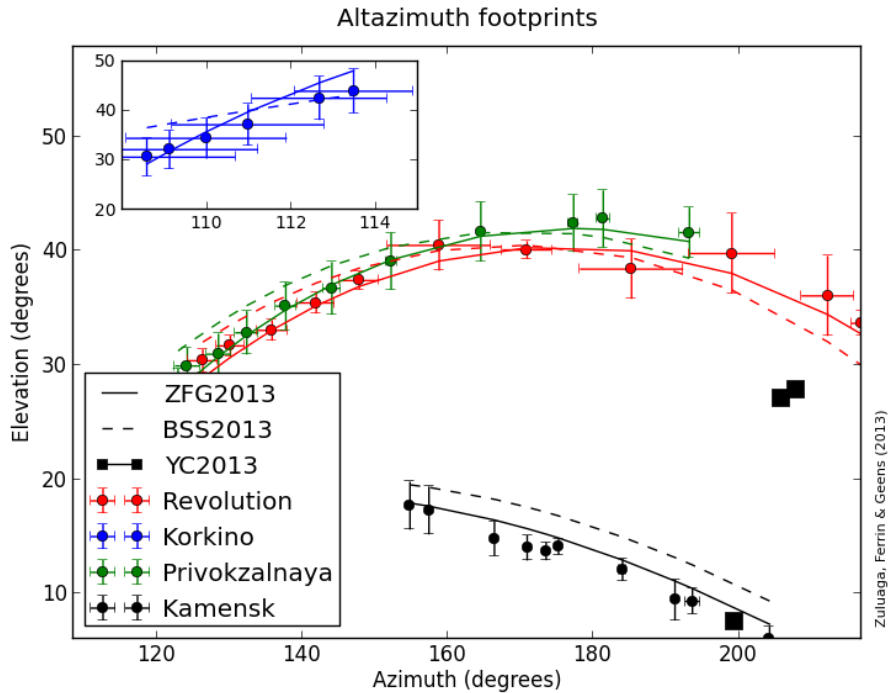

**Figure 2. The altazimuth-footprint triangulation method.** Every observer's location has an elevation vs. altitude signature (dots with error bars). Three signatures pinpoint the orbit very precisely. Notice that time is not involved in this plot. The solid lines are theoretical determinations using the best-fit orbit obtained in this work (ZFG2013). We have also included the theoretical prediction by the orbit solution provided by the team of the Astronomical Institute of the Czech Academy of Sciences (BSS2013, Borovicka et al. 2013) and NASA (YC2013, Yeomans & Chodas, 20013). Our reconstruction is very satisfactory.

Each test trajectory was characterized by 5 parameters: longitude, latitude, altitude, azimuth and elevation as measured in the projected impact point. The best-fit trajectory was calculated by minimizing the least-square statistics with respect to the previous 5 parameters. Given the fact that the measured elevations are derived from a measuring procedure affected by normal errors, and since the number of observations was statistically significant (38 points), we assume that the statistics are distributed according to a Chi-square distribution with 33 degrees of freedom. Using this fact, we compute for each trajectory the one-tailed p-value and use it to compute minimum and maximum limits for the trajectory parameters. The altazimuth-footprint method presented here is trivially applicable to an arbitrary number of locations.

| Projected Impact Site | Best-fit | Minimum | Maximum |
|---|---|---|---|
| Longitude | 59.8703E | +0.051E | -0.043E |
| Latitude | 55.0958N | +0.15N | -0.19N |
| Altitude | 219.0 m | +0.4 m | -0.4 m |
| Azimuth* | 105.0° | +2.2° | -1.7° |
| Elevation* | 15.8° | +0.27° | -0.32° |
| Right Ascension | 324.3° | +1.66° | -1.51° |
| Declination | 4.73° | +1.18° | -1.12° |
| Velocity | 16.7 km/s | +0.65 km/s | -0.68 km/s |
| Height | 68.3 km | +3.62 km | -3.30 km |

* Bolide radiant
** Difference of maximum and minimum with respect to best-fit trajectory in a sample with n=153 trajectories with p-value 0.05<p<0.95.

Table 1. **Properties of the best-fit trajectory for this work**. The trajectory is assumed rectilinear and having a projected point on the ground located at the given latitude and longitude. Velocity was estimated from the Kamensk Uralsky vantage point using the highest observed point in the trajectory.

It should be stressed that elevations and azimuths were computed properly with respect to the average geoid, with equatorial radius $R_E = 6378.2064$ and flattening $f = 1/294.9787$ (Acton, 1996). All the calculations involving reference systems and ephemerides were performed using the NOVAS Package, developed by the U.S. Naval Observatory (Bangert et al. 2011) and the SPICE Toolkit (Acton, 1996).

## 3. Results

After applying the altazimuth-footprint method to our set of vantage points and observations, we obtained the properties of the best-fit trajectory that we summarize in Table 1. According to our triangulation the projected impact site was not even close to Lake Chebarkul, as is assumed in previous reconstruction attempts (Zuluaga & Ferrin, 2013). This implies that if a piece of the parent body is found in the lake it is a fragment that separated during entry. The projected impact site misses by several kilometers the city of Miass, 83 km to the west of Chelyabinsk.

In order to reconstruct the heliocentric orbit of the impactor we took as the *orbit reference point* the highest observed point in the trajectory as registered from the farthest vantage point, i.e. the central square of Kamensk-Uralsky. This particular point in the trajectory is suitable because: 1) It is high enough in the atmosphere (71 km in our best fit trajectory) to ensure that the estimated speed is the least affected by atmospheric drag and 2) the speed can be reliably estimated by direct observations of the fireball performed at this vantage point. Using the available information we estimate that the velocity of the impactor at that height was 16.0-17.4 km/s. This is almost 1-2 km/s below NASA estimate (Yeomans & Choda et al. 2013) and coincides in the upper end with the estimation by the Czech group (Borovicka et al. 2013).

We transform the geodetic state vector at the orbit reference point (position expressed in latitude, longitude and altitude and velocity expressed in terms of speed, azimuth and elevation), first to cartesian planetocentric coordinates in the rotating International Terrestrial Reference System (ITRS), and from there to the Geocentric Celestial Reference System (GCRS). Using the JPL ephemeris for the Earth at the exact time of the impact, we finally obtain the precise state vector of the impactor as referred to the barycenter of the solar system. Although a direct conversion of the state vector to classical elements as referred to the Sun can be performed at this point, we preferred to numerically integrate the trajectory of the body backwards up to 4 years before the impact. For that purpose we used the Mercury integrator (Chambers, 2008), setting precisely the position of the major Solar System planets. We did this to know if small planetary or lunar perturbations could have altered the trajectory previous to the body impact. No significant perturbations were found in this time frame. The orbital elements reported in

| Property | Best-fit orbit | Standard Deviation | Min. | Max. |
|---|---|---|---|---|
| a | 1.266 AU | 0.033 | 1.21 | 1.34 |
| e | 0.435 | 0.018 | 0.40 | 0.47 |
| q | 0.716 AU | 0.005 | 0.71 | 0.72 |
| Q | 1.816 AU | 0.007 | 1.71 | 1.97 |
| i | 2.984° | 0.19 | 2.76 | 3.46 |
| ω | 95.08° | 0.76 | 93.4 | 96.8 |
| Ω | 326.54° | 0.08 | 326.5 | 326.8 |
| M* | 106.85° | - | - | - |

All elements are referred to J2000.0
* Measured at the epoch of impact, jd = 2456338.639279.

**Table 2 | Elements of the best-fit heliocentric orbit**. Standard deviation, minimum and maximum were calculated for a sample of n=50 orbits having p-values of the least-square statistics 0.05<p<0.95.

In table 2 we present the osculant elements measured at 0.01 years before impact (Earth-body distance, $\Delta > 0.05$ AU). In figure 3 we show the orbital elements *a* and *e* as compared with that of three NEA families, the Amors, the Apollos and the Athens. We see that the parent asteroid of the Chelyabinsk event can be classified unambiguously as an Apollo asteroid, confirming preliminary attempts (Zuluaga & Ferrin, 2013). It is also interesting to notice that Tunguska impactor may have been an Apollo (Andreev, 1990). The two impactors have suspisciously similar atmospheric trajectories (impact time, azimuth, elevation and geographic location) (Sekanina, 1998) and hence they may be related. However any detailed analysis of this particular issue is beyond the scope of this paper.

Using the best-fit orbit, we then compute the observational parameters of the parent body a few days before impact (see figure 4). To estimate the brightness of the we perform a backward integration of the orbit with a small step size (6 hours). We assumed a geometric albedo $p_V = 0.28 \pm 0.13$ which is the mean value of 9 Apollo asteroids (Veeder, 1989), and a diameter of 18 m (Yeomans & Chodas, 2013). Three phase laws were selected (see e.g. Delahod et al. 2001). These values imply that the object had an absolute magnitude H = 25.8. Only a few hours before impact does the apparent magnitude rise to levels detectable by current surveys. However, its angular distance to the sun is lower than the search limit of these same surveys. Having an absolute magnitude H>22, the object has not been classified as a Potentially Hazardous Asteroid, PHA. The object should actually be classified as an *HA* or a *Harmful Asteroid,* of which it is the second member after the Tunguska event impactor. Since the current number of officially known PHAs, i.e. $\Delta_{min}<0.05$ AU and H<22, is 1381, an extrapolation of the absolute magnitude cumulative distribution from H=22 to H=25.8 gives an estimate of 1300±50 additional potentially hazardous asteroids. These objects are not capable of global damage but are still capable of local damage. With this result the number of PHAs larger than the Chelyabinsk event impactor is raised from the current 1381 to 2680±70 bodies, almost double the previous value.

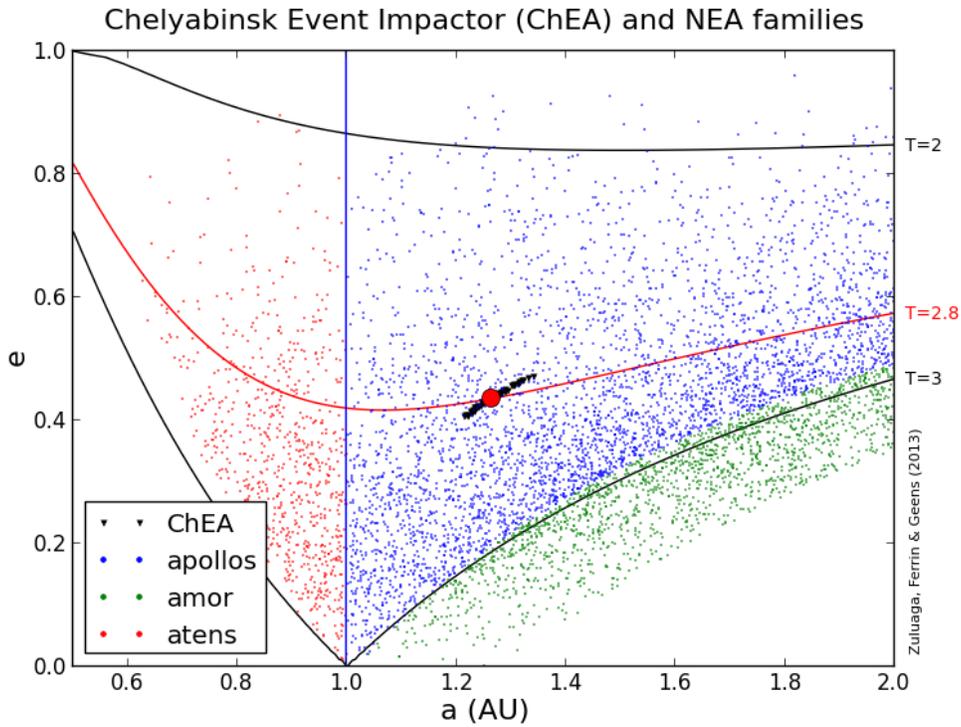

**Figure 3. Orbital classification of the Chelyabinsk event impactor**. Big red point indicates the position of the best-fit orbit elements. Black diamonds show n=50 orbits with p-value of the altazimuth-footprint least square 0.05<p<0.95. The object lies in the region of the Apollo asteroids and has a Tisserand invariant of 2.8.

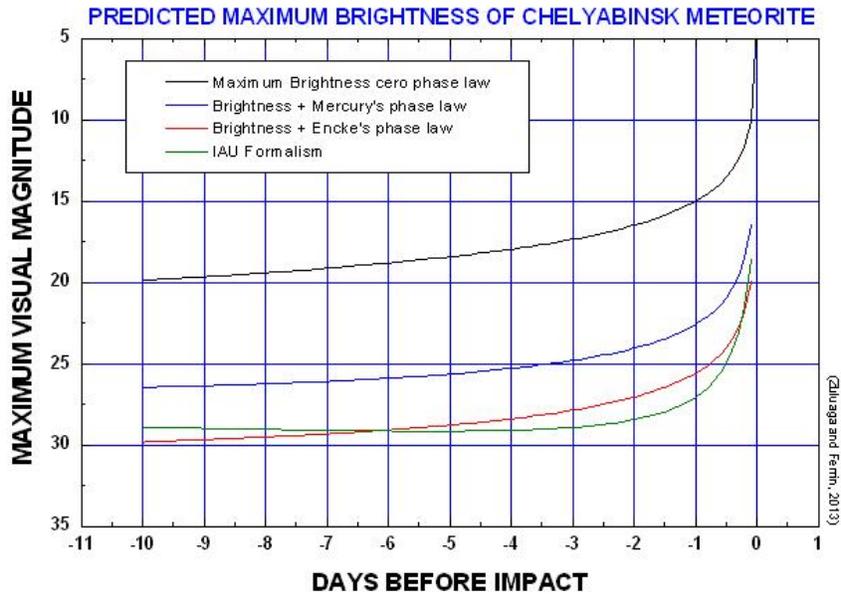

**Figure 4. Calculation of the asteroid brightness in the days before impact**. Just few hours before the impact the parent body could be potentially observed by NEA surveys. However since its angular distance to the sun was probably less than 15º it was out of the scope of the present visual surveys.

## 5. Conclussions

There are several lessons we can learn from this work. (1) To confirm the fundamental role that active enthusiasts, a.k.a. *citizen astronomers*, can play in scientific research, especially in cases when unexpected events occur. (2) We are missing half of the PHAs by our own definition of what a PHA is. (3) Although objects smaller than 100 m cannot produce global damage they can still produce significant local damage. (4) A simple calculation shows that if the impactor had been delayed by about 3.5-4 minutes, the impact would have taken place over central Europe, were the damage could have been much greater. (5) The object approached Earth from the Sun side, which is not covered by current optical surveys. This side is totally unshielded.

## Acknowledgements

The data contributing to the successful reconstruction of the orbit were taken by people in Chelyabinsk, Korkino and Kamensk-Uralsky. We thank these contributors for submitting their footage to public access. We also want to thank those citizen scientists who alerted us to useful videos or pinpointed their location information in the comments to the original blog post on Ogle Earth (http://ogleearth.com) in the days after the event. Special thanks to *SebastienP, liilliil, Sean Mac, Robin Whittle, Kuuuurija, Serge, Latuha Valeryi, ssvilponis, Dmitry DD, comeT, Sirius, g1smd, Steve* and *Gary* (Online handles are self-reported, information was independently verified). J.Z. and I.F. acknowledge support from the CODI/UdeA.

## Appendix A. The videos

All the footage used in this work is publicly available from videos on YouTube (verified March 4, 2013). The location of the vantage points where the videos were recorded and the YouTube URL for each of them are listed in the table below:

| Vantage point | Longitud, Latitude | Altitude | URL |
| --- | --- | --- | --- |
| Revolution Square (Chelyabinsk) | 61.40249°E, 55.16080°N | 230 m | https://www.youtube.com/v/bXifSi2K278 |
| Privokzalnaya Square (Chelyabinsk) | 61.41421°E, 55.14392°N | 238 m | https://www.youtube.com/v/Qin41lP9r2U |
| Market Place (Korkino) | 61.39956°E, 54.89093°N | 241 m | https://www.youtube.com/v/odKjwrjIM-k |
| Central Square (Kamensk Uralsky) | 61.91858°E, 56.41500°N | 169 m | https://www.youtube.com/v/iCawTYPtehk |

**Table A1**. Location of the vantages points and link to the videos used in the reconstruction.

The authors have preserved backup copies of the videos, in case they are removed from YouTube. No public access to these backups is available. In the event the videos are no longer available on YouTube, researchers interested in the material should contact jzuluaga@fisica.udea.edu.co for a copy.

A complete Google Earth KMZ file showing the trajectory and referencing the vantage points used in the reconstruction. It can be download from this URL:

http://urania.udea.edu.co/sitios/facom/pages/chelyabinsk-meteoroid.rs/files/chelyabinsk-meteoroidfkpdk/ZFG2013-TrajectoryVantagePointsVideos.kmz

The positions of other locations with links to videos recorded at those locations are referenced in the same file.

Other supplementary information and updates are available here:
 http://astronomia.udea.edu.co/chelyabinsk-meteoroid.